\documentclass[]{tMOP2e}

\newcommand{\RM}{\mathbb{R}}
\newcommand{\SM}{\mathbb{S}}

\newcommand{\beq}{\begin{equation}}
\newcommand{\eeq}{\end{equation}}
\newcommand{\barr}{\begin{eqnarray}}
\newcommand{\earr}{\end{eqnarray}}

\usepackage{bm}
\usepackage{color}

\begin{document}

\doi{10.1080/0950034YYxxxxxxxx}
 \issn{1362-3044}
\issnp{0950-0340} \jvol{00} \jnum{00} \jyear{2008} \jmonth{10 January}

\markboth{Taylor \& Francis and I.T. Consultant}{Journal of Modern Optics}

%\articletype{GUIDE}

\title{On the inversion of the Radon transform: standard vs $M^2$ approach}

\author{Paolo Facchi$^{a,b}$,
%$^{\ast}$\thanks{$^\ast$Corresponding author. Email: paolo.facchi@ba.infn.it \vspace{6pt}}
Marilena Ligab\`o$^{a}$ and Saverio Pascazio$^{c,b}$\\\vspace{6pt}
$^{a}${\em{Dipartimento di Matematica, Universit\`a di Bari,
        I-70125  Bari, Italy}} \\
$^{b}${\em{INFN, Sezione di Bari, I-70126 Bari, Italy}} \\
$^{c}${\em{Dipartimento di Fisica, Universit\`a di Bari,
I-70126  Bari, Italy}}\\\vspace{6pt}\received{May 2009} }

\maketitle

\begin{abstract}
We compare the Radon transform in its standard and symplectic formulations and argue that the inversion of the latter can be performed more efficiently.
\bigskip
\end{abstract}

\bigskip

\section{Introduction}
\label{sec-introd}

The Radon transform \cite{Rad1917} is a key mathematical tool in
tomography. Its inverse enables one to reconstruct a function if
some of its integrals are known. The whole subject has been recently
revived by quantum mechanical applications. The possibility of
reconstructing the tomographic map of the Wigner quasidistribution
function
\cite{Wig32,Moyal,Hillary84} associated with a given quantum state
\cite{Ber-Ber,Vog-Ris,Mancini95} has motivated experiments
\cite{SBRF93,torino,konst},
triggered novel proposals \cite{reconstruct06} and boosted
innovative theoretical techniques \cite{theory}. Applications are
widespread and diverse. The entire field, driven by a blizzard of
technical advances, is attracting increasing attention and is
growing at a lively pace. Good reviews on the subject can be found
in \cite{Jardabook}.

The Radon transform was originally introduced as an integral
transform defined over submanifolds of $\mathbb{R}^n$, that may be
viewed as a ``configuration space." However, if $n$ is even, one may
think of $\mathbb{R}^n$ as a phase space and consider the integrals
over its Lagrangian submanifolds. One may then associate the
tomographic map with the symplectic transform on the phase space
\cite{dariano96}. In this context, motion is instrumental for the
identification of the phase space and its Lagrangian variables: the
Hamilton equations do not appear in the definition of the Radon
transform and this interpretation differs from the original one.
Nevertheless, the approach is prolific and enables one to identify
different types of tomograms \cite{PRA07}, extend tomography to
curved surfaces \cite{PRA08} and consider more general problems and
applications. This is in line with previous hystorical developments,
by Radon himself \cite{Rad1917}, John \cite{John}, Helgason
\cite{Helgason} and Strichartz \cite{Strichartz}, and paves the
way towards so far unhearted quantum mechanical applications.

In this article we shall compare the standard Radon approach with
that based on the afore-mentioned symplectic identification. We
shall argue that, although mathematically equivalent, they may
differ in practice. In particular, the inversion may be far from
trivial and may turn out to be simpler in the symplectic framework.

\section{Symplectic tomography}
\label{sec-sympltomog}

Let us focus on the 2-dimensional case for the sake of concreteness.
The Radon transform, in its original formulation, solves the
following problem: reconstruct a function of two variables, say
$f(p,q)$, if its integrals over arbitrary lines are given. The Radon
transform (or homodyne tomogram) reads
\begin{equation}
f^{\sharp}(\vartheta,X)=\int_{\RM^2} f(q,p)
\delta(X-q\cos\vartheta-p\sin\vartheta)\; dq dp ,
\label{eq:fsharpexpl}
\end{equation}
where $\delta$ is the Dirac function, $\vartheta \in [0,2 \pi)$,
$\omega=(\cos\vartheta,\sin\vartheta)\in S$ (the unit sphere in 1D)
and $X \in \RM$. In order to obtain a symplectic formulation, a
central observation is the following: it is possible to express the
Radon transform in affine language (the so-called tomographic map)
\cite{Rad1917,Gelf} and write
\begin{equation}
f^{M^2}(\mu,\nu,X)=\int_{\RM^2}f(q,p) \delta(X-\mu q-\nu
p)\; dq dp ,
\label{eq:fm2expl}
\end{equation}
where $\mu,\nu,X \in \RM$, $\mu^2+\nu^2>0$. We have named ``$M^2$"
the tomographic map (\ref{eq:fm2expl}) after Man'ko and Marmo, who
gave seminal contribution towards its significance
\cite{MarmoPhysScr,MarmoJPA,MarmoPL,Marmoopen}. Clearly
\begin{equation}
f^{\sharp}(\vartheta,X)=f^{M^2}(\cos\vartheta,\sin\vartheta,X).
\label{eq:ff2}
\end{equation}
Consider now a particle moving on the line $q \in \mathbb{R}$ and a
function $f(q,p)$ on its phase space $(q,p) \in \mathbb{R}^2$. Since
\begin{equation}
\left(
  \begin{array}{cc}
    \mu & \nu \\
  \end{array}
\right) \left( \begin{array}{cc}
    1 & 0 \\
    0 & 1 \\
  \end{array}
\right) \left(
  \begin{array}{c}
    q \\
    p \\
  \end{array}
\right) = \mu q + \nu p
=
\left(
  \begin{array}{cc}
    -\nu & \mu \\
  \end{array}
\right) \left(
  \begin{array}{cc}
    0 & -1 \\
    1 & 0 \\
  \end{array}
\right) \left(
  \begin{array}{c}
    q \\
    p \\
  \end{array}
\right) ,
\end{equation}
the argument in the Dirac delta function in Eq.\ (\ref{eq:fm2expl})
may be considered either as a Euclidean product or as a symplectic
product. The two interpretations are completely equivalent and one
can equivalently solve the inversion problem by using the Euclidean
or symplectic Fourier transform. We shall see in the next section
that the two procedures can vastly differ in complexity.

Note that the Radon transform is defined in an equivalent way by
\begin{equation}
f^{\sharp}(\vartheta,X) =\int_{\mathbb{R}} f(s\cos\vartheta
-X\sin \vartheta,s\sin \vartheta + X\cos \vartheta) ds .
\end{equation}
The inversion formula, as given by Radon, amounts to consider first
the average value of $f^{\sharp}$ on all lines tangent to the circle of
center $(q,p)$ and radius $r$, namely,
\begin{equation}
F_{(q,p)}(r)=\frac{1}{2\pi}\int_0^{2\pi} f^{\sharp}(\vartheta, q \cos \vartheta + p\sin \vartheta +
r) d \vartheta
\label{eq:circlemean}
\end{equation}
and then compute
\begin{equation}
f(q,p)=-\frac{1}{\pi}\lim_{\varepsilon \downarrow 0}\int_\varepsilon^{+\infty} F'_ {(q,p)} (r) \frac{dr}{r},
\label{eq:Randoninversion}
\end{equation}
where $F'_ {(q,p)}(r)$ denotes the derivative with respect to $r$.
The Radon transform maps a (suitable) function on the plane into a
function on the cylinder. Some conditions that guarantee the
invertibility and continuity of the map were studied by Radon
himself \cite{Rad1917}, John \cite{John}, Helgason \cite{Helgason}
and Strichartz \cite{Strichartz}.

On the other hand, the inverse transform of (\ref{eq:fm2expl}) reads
\cite{MarmoJPA,MarmoPhysScr}
\begin{eqnarray}
f(q,p) &=&  \int_{\mathbb{R}^3} f^{M^2}(\mu,\nu,X) e^{i(X-\mu q-\nu
p)} \frac{dXd\mu d\nu}{(2\pi)^2} . \quad \label{eq:invgen}
\end{eqnarray}

\section{The inverse transform: an explicit example}

We now compare the inversions
(\ref{eq:circlemean})-(\ref{eq:Randoninversion}) and
(\ref{eq:invgen}) by looking at a very simple example: the ground
state of a one dimensional quantum harmonic oscillator,
\begin{equation}
f(q,p)=\frac{\alpha^2}{\pi}e^{-\alpha^2(q^2+p^2)}.
\label{eq:fqp}
\end{equation}
Its $M^2$-transform reads
\begin{eqnarray}
f^{M^2}(\mu,\nu,X) & = & \frac{\alpha^2}{\pi}\int_{\RM^2}
e^{-\alpha^2(q^2+p^2)} \delta(X-\mu q-\nu p)\; dq dp \nonumber \\
                   & = & \frac{\alpha^2}{|\mu|\pi}\int_{\RM}e^{-\alpha^2\left[\left( \frac{X-\nu
                   p}{\mu}\right)^2+p^2\right]}\; dp \nonumber\\
                   & = & \frac{\alpha^2}{|\mu|\pi}\int_{\RM} e^{-\frac{\alpha^2}{\mu^2}\left[
                   \left( \sqrt{\mu^2+\nu^2}p-\frac{\nu X}{\sqrt{\mu^2+\nu^2}}\right)^2+ \frac{\mu^2X^2}{\mu^2+X^2}
                   \right]}\; dp \nonumber\\
                   & = &
                   \frac{\alpha}{\sqrt{\pi(\mu^2+\nu^2)}}e^{-\alpha^2
                   \frac{X^2}{\mu^2+\nu^2}},
\end{eqnarray}
which is a Gaussian with respect to $X$, but has a nontrivial
dependence on $\mu$ and $\nu$. On the other hand, by making use of
(\ref{eq:ff2}), one gets the Radon transform
\begin{equation}
f^{\sharp}(\vartheta,X)=f^{M^2}(\cos \vartheta, \sin
\vartheta,X)=\frac{\alpha}{\sqrt{\pi}}e^{-\alpha^2 X^2},
\label{eq:fqp1}
\end{equation}
which is simply a Gaussian, independent of the angle $\vartheta$,
due to symmetry.

Let us tackle the inversion problem. We start from the inverse $M^2$ transform,
which is easily solved in a few lines:
\begin{eqnarray}
f(q,p) & = & \frac{1}{(2\pi)^2}\int_{\RM^3}
f^{M^2}(\mu, \nu, X)e^{i(X-\mu q-\nu p)} \; dX d \mu d\nu \nonumber \\
       & = & \frac{\alpha}{(2\pi)^2 \sqrt{\pi(\mu^2+\nu^2)}}\int_{\RM^3} e^{-\frac{\alpha^2}{\mu^2+\nu^2}\left(X^2-iX \frac{\mu^2+\nu^2}{\alpha^2}
       \right)}e^{-i(\mu q+\nu p)} \; dX d \mu d\nu \nonumber \\
       & = & \frac{1}{(2\pi)^2}\int_{\RM^2} e^{-\frac{\mu^2+\nu^2}{4
       \alpha^2}}e^{-i(\mu q+\nu p)}\; d\mu d\nu \nonumber \\
       & = & \frac{\alpha}{\pi} e^{-(q^2+p^2) \alpha^2} \label{eq:ex1} .
\end{eqnarray}

Let us now endeavour to invert the Radon transform. It would be
tempting to leave this as an exercise for the reader, but we will
sketch the main steps of the derivation. From (\ref{eq:circlemean})
we get
\begin{equation}
F'_{(q,p)}(r)= -\frac{\alpha^3}{\pi\sqrt{\pi}} \int_0^{2\pi} (q\cos\vartheta+ p\sin\vartheta +r )
e^{-\alpha^2 (q\cos\vartheta+ p\sin\vartheta +r )^2} d\vartheta
\end{equation}
and thus $f(q,p)=\lim_{\varepsilon\downarrow0} f_{\varepsilon}(q,p)$, where
\begin{equation}
f_\varepsilon(q,p)%=-\frac{1}{\pi}\lim_{\varepsilon \downarrow 0}\int_\varepsilon^{+\infty} F'_ {(q,p)} (r) \frac{dr}{r}
= \frac{\alpha^3}{\pi^2\sqrt{\pi}}  \int_\varepsilon^{+\infty} dr  \int_0^{2\pi} d\vartheta
\frac{(q\cos\vartheta+ p\sin\vartheta +r )}{r}
e^{-\alpha^2 (q\cos\vartheta+ p\sin\vartheta +r )^2} .
\end{equation}
Already in the very simple case of a Gaussian function with a
Gaussian Radon transform the above inversion formula is not easy to
manage. First introduce a step function, $\theta(r)=1$ only if
$r>0$, and change the period of the angle integration
\begin{equation}
f_\varepsilon(q,p)
= \frac{\alpha^3}{\pi^2\sqrt{\pi}}   \int_{\mathbb{R}} dr \int_{-\pi}^{\pi} d\vartheta
\frac{\theta(r-\varepsilon)}{r}
(r- q\cos\vartheta+  - p\sin\vartheta  )
e^{-\alpha^2 (r- q\cos\vartheta - p\sin\vartheta  )^2} .
\end{equation}
Then change variables $z=r- q\cos\vartheta - p\sin\vartheta$ and $t=\tan (\vartheta/2)$
\begin{eqnarray}
f_\varepsilon(q,p)
&=& \frac{2\alpha^3}{\pi^2\sqrt{\pi}}   \int_{\mathbb{R^2}}
\theta\left(\frac{t^2(z-q-\varepsilon)+2 p t +(z+q-\varepsilon) }{1+t^2}\right)
\nonumber\\
& & \qquad\qquad  \times \frac{z e^{-\alpha^2 z^2}}{t^2(z-q)+2 p t +(z+q)}\, dt\, dz\,
 .
\end{eqnarray}
Now look at the region where the argument of the theta function is positive. One gets
two roots
\begin{equation}
t_{1,2}= \frac{-p\pm \sqrt{p^2+q^2-z^2-2\varepsilon z -\varepsilon^2}}{z-q-\varepsilon}
\end{equation}
whose discriminant is negative for $z\notin [-\sqrt{p^2+q^2}+\varepsilon , +\varepsilon\sqrt{p^2+q^2}+\varepsilon]$. Therefore,
\begin{equation}
f_\varepsilon(q,p)=I_1(\varepsilon; q,p)+I_2(\varepsilon; q,p),
\end{equation}
where
\begin{equation}
I_1(\varepsilon, q,p)= \frac{2\alpha^3}{\pi^2\sqrt{\pi}}   \int_{\sqrt{p^2+q^2}+\varepsilon}^{+\infty}
dz\,  z e^{-\alpha^2 z^2} \int_{-\infty}^{+\infty } dt\,  \frac{1}{t^2(z-q)+2 p t +(z+q)}
\end{equation}
and
\begin{eqnarray}
I_2(\varepsilon, q,p)&=& \frac{2\alpha^3}{\pi^2\sqrt{\pi}}   \int_{-\sqrt{p^2+q^2}+\varepsilon}^{\sqrt{p^2+q^2}+\varepsilon}
dz \int_{-\infty}^{+\infty } dt\,  \theta\left(\frac{t^2(z-q-\varepsilon)+2 p t +(z+q-\varepsilon) }{1+t^2}\right)
\nonumber\\
& & \qquad\qquad \qquad\qquad\qquad\qquad \times \frac{z e^{-\alpha^2 z^2}}{t^2(z-q)+2 p t +(z+q)}
 .
\end{eqnarray}
Let us evaluate $I_1(\varepsilon, q,p)$. The integration over $t$ yields $\pi/ \sqrt{z^2-q^2-p^2}$ and thus
\begin{equation}
I_1(\varepsilon, q,p)= \frac{2\alpha^3}{\pi\sqrt{\pi}}   \int_{\sqrt{p^2+q^2}+\varepsilon}^{+\infty}
 \frac{z e^{-\alpha^2 z^2}}{\sqrt{z^2-q^2-p^2}}\,   dz.
\end{equation}
An  integration by part gives
\begin{eqnarray}
I_1(\varepsilon, q,p) &=& \frac{4\alpha^5}{\pi\sqrt{\pi}}   \int_{\sqrt{p^2+q^2}+\varepsilon}^{+\infty}
 z \sqrt{z^2-q^2-p^2} e^{-\alpha^2 z^2}\,   dz + O(\sqrt{\varepsilon})
\nonumber\\
&=& \frac{4\alpha^5}{\pi\sqrt{\pi}}   \int_{(2\varepsilon\sqrt{p^2+q^2}+\varepsilon^2)^{1/2}}^{+\infty}
 y^2 e^{-\alpha^2 (y^2+q^2+p^2)}\,   dy + O(\sqrt{\varepsilon})
\nonumber\\
&=& \frac{2\alpha^5}{\pi\sqrt{\pi}}\,  e^{-\alpha^2 (q^2+p^2)}  \int_{-\infty}^{+\infty}
 y^2 e^{-\alpha^2 y^2}\,   dy + O(\sqrt{\varepsilon})
 ,
\end{eqnarray}
where $y^2=z^2-q^2-p^2$. Since the Gaussian integral equals $\sqrt{\pi}/2 \alpha^3$, we finally get
\begin{equation}
\lim_{\varepsilon\downarrow0} I_1(\varepsilon, q,p)= f(q,p).
\end{equation}
Therefore, it remains to prove that $I_2(\varepsilon, q,p)$ vanishes for $\varepsilon\to0$.
We will leave it as a very instructive exercise.

Notice that the Radon transform can also be inverted by using the
following alternative formula due to Helgason \cite{Helgason}, which
is suitable for generalizations to symmetric homogeneous spaces
\begin{eqnarray}
f(q,p)  =  \frac{1}{4 \pi}
(-\Delta)^{1/2}\int_{0}^{2\pi}f^{\sharp}(\vartheta,q \cos
\vartheta+ p \sin \vartheta) \; d\vartheta .
%\nonumber \\
 %      & = & \frac{\alpha}{4 \pi \sqrt{\pi}} \left( -\frac{\partial^2}{\partial q^2}-\frac{\partial^2}{\partial p^2}
  %     \right)^{1/2}\int_{0}^{2\pi} e^{-\alpha^2(q \cos \vartheta + p \sin
   %    \vartheta)^2}\; d \vartheta. \label{eq:ex2}
\end{eqnarray}
Here the fractional Laplacian
\begin{equation}
(-\Delta)^{1/2}= \left( -\frac{\partial^2}{\partial q^2}-\frac{\partial^2}{\partial p^2} \right)^{1/2}
\end{equation}
is defined by a Fourier transform
\begin{eqnarray}
(-\Delta)^{1/2} g(q,p) = \int_{\mathbb{R}^2} (k_1+k_2)^{1/2} \hat{g}(k_1,k_2) e^{i(q k_1+p k_2)} \frac{dk_1 dk_2}{2\pi},
\end{eqnarray}
where
\begin{equation}
\hat{g}(k_1,k_2) = \int_{\mathbb{R}^2} g(q,p) e^{-i(q k_1+p k_2)} \frac{dq\, dp}{2\pi}.
\end{equation}
In our case we would have to compute
\begin{eqnarray}
\frac{\alpha}{4 \pi \sqrt{\pi}} \left( -\frac{\partial^2}{\partial q^2}-\frac{\partial^2}{\partial p^2}
       \right)^{1/2}\int_{0}^{2\pi} e^{-\alpha^2(q \cos \vartheta + p \sin
       \vartheta)^2}\; d \vartheta, \label{eq:ex2}
\end{eqnarray}
a task even more difficult than the previous one.

\section{Extension to $n$ dimensions and discussion}

The definitions and conclusions of the previous sections can be easily extended
to $n$ dimensions.
The Radon transform of a function $f$ of the $n$-dimensional vector $x\in \RM^n$
reads
\begin{equation}
f^{\sharp}(\omega,X)=\int_{\RM^n}f(x)
\delta(X-\langle\omega,x\rangle)\; dx ,
\label{eq:fsharp}
\end{equation}
where $\omega \in \SM^{n-1}$, the $(n-1)$-dimensional sphere,
$\langle \cdot , \cdot \rangle$ denotes scalar product and $X \in
\RM$. The $M^2$ transform is
\begin{equation}
f^{M^2}(\mu,X)=\int_{\RM^n}f(x) \delta(X-\langle\mu,x\rangle)\; dx ,
\label{eq:fm2}
\end{equation}
where $\mu \in \RM^n$ and $X \in \RM$. Obviously, from
$f^{M^2}(\mu,X)$ one can immediately recover $f^{\sharp}(\omega,X)$
by setting $\mu=\omega\in \SM^{n-1}$:
\begin{eqnarray}
f^{\sharp}(\omega,X)=f^{M^2}(\omega,X).
\label{eq:ff0}
\end{eqnarray}
However, notice that, although $f^{\sharp}$ is the restriction of
$f^{M^2}$ on the unit sphere $\SM^{n-1}$, there is actually a
bijection between the two transforms. Therefore, they carry exactly
the same information. Indeed, since the Dirac distribution is
positive homogeneous of degree $-1$, i.e.\ $\delta(\alpha x)=
|\alpha|^{-1} \delta(x)$, for every $\alpha\neq 0$, one gets from
Eq.\ (\ref{eq:fm2})
%\begin{equation}
%f^{M^2}(\alpha \mu, \alpha X)= |\alpha|^{-1} f^{M^2}(\mu, X),
%\end{equation}
%for every $\alpha \neq 0$. Therefore, by taking $\alpha = 1/|\mu|$ one gets
\begin{eqnarray}
f^{M^2}(\mu,X)=\frac{1}{|\mu|}f^{M^2}\left( \frac{\mu}{|\mu|},
\frac{X}{|\mu|} \right),
\end{eqnarray}
for $\mu \neq 0$. In words, the tomogram $f^{M^2}(\mu,X)$ at a
generic point $\mu \in \RM^n$ is completely determined by the
tomogram at $\mu/|\mu| \in \SM^{n-1}$. But the latter is nothing but
the Radon transform, by Eq.\ (\ref{eq:ff0}). Therefore we get the
bijection
\begin{eqnarray}
f^{\sharp}(\omega,X)&=&f^{M^2}(\omega,X),
\label{eq:ff} \\
f^{M^2}(\mu,X)&=&\frac{1}{|\mu|}f^{\sharp}\left( \frac{\mu}{|\mu|},
\frac{X}{|\mu|} \right) \qquad (\mu \neq 0).
\label{eq:ff33}
\end{eqnarray}
Notice also that at the origin $\mu=0$ the $M^2$ transform
\begin{equation}
f^{M^2}(0,X)=\delta(X) \int_{\RM^n}f(x) \; dx ,
\end{equation}
depends only on the total mass.

The inversion formulae for the transforms (\ref{eq:fsharp}) and
(\ref{eq:fm2}) read
\begin{equation}
f(x)  =  \frac{1}{2^n \pi^{n-1}}
(-\Delta)^{(n-1)/2}\int_{\SM^{n-1}} f^{\sharp}(\omega,\langle\omega,x\rangle) \; d\omega
\label{eq:geninvrad}
\end{equation}
and
\begin{equation}
f(x) =  \int_{\mathbb{R}^{n+1}} f^{M^2}(\mu ,X)\,
e^{i(X-\langle\mu,x\rangle)}\, \frac{dX d\mu}{(2\pi)^n} ,
\label{eq:geninvsym}
\end{equation}
respectively. While formula (\ref{eq:geninvsym}), which is nothing
but a Fourier transform, is quite easy to handle, the inversion
formula (\ref{eq:geninvrad}) is in general very hard to tackle,
especially for even $n$, due to the presence of a fractional
Laplacian. Therefore, from a practical point of view our message is
the following: in order to invert the Radon transform
(\ref{eq:ff0}), dilate it by (\ref{eq:ff33}) into the $M^2$
transform and then use the Fourier inversion formula
(\ref{eq:geninvsym}). This simple trick enables one to avoid long
and tedious calculations.

\section*{Acknowledgements}
This work was partially supported by the EU through the Integrated
Project EuroSQIP. We thank V. Man'ko and G. Marmo for many
discussions on the meaning of the Radon transform and G. Florio for
a bright suggestion.

\end{document}